\documentclass{aa} 

\usepackage{graphicx}  
\usepackage{txfonts}  
\usepackage{natbib}
\bibpunct{(}{)}{;}{a}{}{,}

\begin{document}  
 
\title{Importance of fingering convection for accreting 
white dwarfs in the framework of full evolutionary calculations: the case
of the hydrogen-rich white dwarfs GD133 and G29-38}
  
  \author{ F. C. Wachlin\inst{1},
           G. Vauclair\inst{2,3},
           S. Vauclair\inst{2,3},
           \and
           L. G. Althaus\inst{1}
           }
\institute{Instituto de Astrof\'{\i}sica de La Plata (UNLP - CONICET). 
           Facultad de Ciencias Astron\'omicas y Geof\'{\i}sicas. 
           Universidad Nacional de La Plata, Argentina\\
           \email{fcw@fcaglp.unlp.edu.ar}     
           \and
           Universit\'e de Toulouse, UPS-OMP, IRAP, France
           \and
           CNRS, IRAP, 14 avenue Edouard Belin, 31400 Toulouse, France\\
           }
\date{\today}

\abstract
{A large fraction of white dwarf stars show photospheric chemical composition polluted by heavy elements accreted from a debris disk. Such debris disks result
 from the tidal disruption of rocky planetesimals which had survived to whole stellar evolution from the main sequence to the final white dwarf stage. 
Determining the accretion rate of this material is an important 
step towards estimating the mass of the planetesimals and towards understanding the ultimate fate of the planetary systems.}
{The accretion of heavy material with a mean molecular weight, $\mu$, higher than the mean molecular weight of the white dwarf outer layers, induces a double-diffusive 
instability producing the fingering convection and an extra-mixing. As a result, the accreted material is diluted deep into the star. We explore the effect of this 
extra-mixing on the abundance evolution of Mg, O, Ca, Fe and Si in the cases of the two well studied polluted DAZ white dwarfs: GD~133 and G~29-38.}
{We performed numerical simulations of the accretion of material having a chemical composition similar to the bulk Earth composition. We assumed a continuous and uniform 
 accretion and considered a range of accretion rates from $10^{4}$~g/s to $10^{10}$~g/s. Two cases are simulated, one using the standard mixing length theory (MLT) 
and one including the double-diffusive instability (fingering convection).}
{The double-diffusive instability develops on a very short time scale. The surface abundance rapidly reaches a stationary value while the depth of the zone mixed by 
the fingering convection increases. In the case of GD~133, the accretion rate needed to reproduce the observed abundances exceeds by more than 2 orders of magnitude 
the rate estimated by neglecting the fingering convection. In the case of G~29-38 the needed accretion rate is increased by approximately 1.7 dex.}
{Our numerical simulations of the accretion of heavy elements 
on the hydrogen-rich white dwarf GD~133 and G~29-38 show that 
fingering convection is an efficient mechanism to mix the 
accreted material deeply. 
We find that when fingering convection is taken into account, 
accretion rates higher by 1.7 to 2 dex than those inferred from 
the standard MLT are needed in order to reproduce the abundances 
observed in G 29-38 and GD 133.
}

  
\titlerunning{The role of fingering convection in accreting hydrogen-rich
white dwarfs}
  
\authorrunning{Wachlin et al.}  

\maketitle 

\section{Introduction}  
\label{intro}
\vskip2cm

Since the early discovery of an infra-red excess from the  
hydrogen-rich white dwarf star G29-38  \citep{1987Natur.330..138Z},
the prototype of the DAZ class, it has become evident that a large 
fraction of white dwarfs are surrounded by debris disk from which heavy 
elements accrete at the white dwarf surfaces 
and pollute their chemical composition. 
These debris disks result from the disruption by tidal effects of 
planetesimal orbiting the white dwarfs \citep{2003ApJ...584L..91J}.


The fraction of WDs surrounded by
such debris disks amounts to 56\% in 
the warm DA white dwarfs sample with effective temperature 17000~K $\leq$Teff$\leq$ 27000~K 
analyzed by \cite{2014A&A...566A..34K}.
\cite{2010ApJ...722..725Z} estimate this fraction 
to be $\approx$ 1/3 for the DB white dwarfs in the effective temperature range 
13500~K $\leq$Teff$\leq$ 19500~K.  

The planetesimals have to come close to, or cross through, the tidal radius of 
the white dwarf for being subsequently disintegrated. To achieve this condition, 
their original, presumably circular orbits
must be perturbed by gravitational 
interactions with massive(s) planet(s), of either Jovian or Neptunian sizes, to evolve 
toward elliptical orbits so as to pass close enough to the white dwarf tidal radius. 
A number of scenarii of planetary systems evolution through the giant and asymptotic 
giant branches have demonstrated that massive planets may survive these phases 
of the star evolution and be able to perturb the planetesimal orbits to large eccentricity 
\citep[e.g.][]{
2002ApJ...572..556D, 2012ApJ...747..148D, 2013EPJWC..4706008M,
2013MNRAS.431.1686V, 2014MNRAS.445.2244V, 2014MNRAS.445.2794V,
2015MNRAS.451.2814V, 2015MNRAS.451.3453V, 2015MNRAS.452.1945V,
2016RSOS....3.0571V, 2016MNRAS.458.3942V, 2014MNRAS.439.2442F}.

The recent discovery of planetesimal fragments transiting 
in front of the white dwarf WD 1145+017 is an evidence that 
disintegrations of planetesimals may indeed occur in such circumstances
\citep{2015Natur.526..546V, 
2015arXiv151006434C, 
2016ApJ...818L...7G, 
2016ApJ...816L..22X}.

The analysis of the chemical composition of the polluted 
white dwarf atmospheres shows that the disrupted planetesimals 
are rocky bodies similar to the asteroids in the solar system. 
Their composition is mostly similar to the Earth bulk composition
\citep{2011ApJ...732...90M, 2014ApJ...783...79X, 2014AREPS..42...45J}.

The determination of the chemical composition of polluted white dwarfs offers a 
unique opportunity to scrutinize the composition of the disrupted planetesimals 
orbiting extrasolar planetary systems. However, in order to relate the chemical 
composition of the disrupted bodies to the composition observed at the white dwarf 
surface it is necessary to correctly describe the various processes which affect the material 
once accreted at the white dwarf surface. If there is a convective zone in the stellar outer layers, the accreted elements are first mixed throughout this zone. 
If not, they first accumulate in the photosphere. In both cases, they are 
further mixed in the extra-mixing zone that is triggered by the fingering convection, a 
process that we discuss in the present paper. In addition they settle 
downwards by
gravitational settling through the bottom of the mixed zone.
Previous estimates were based on the assumption that the 
accreted material diffuses downwards by gravitational settling only
\citep[e.g.][]{2012MNRAS.424..464F, 2014A&A...566A..34K}. 

However, the double-diffusive instability resulting from the inverse $\mu$-gradient
induced by the accretion of heavy elements
 is much more efficient that the graviational setling and has to be taken
into account. This kind of instability is well known in 
oceanography where it is refereed to as the thermohaline 
convection. In the oceanographic context the instability 
occurs in situations where 
warm salted water lies on top of cooler fresh water.
It leads to the so-called salt fingers. 
The same instability has been found to occur in various 
astrophysical situations where an inverse $\mu$ gradient develops 
in a thermally stable layer 
\citep{
2004ApJ...605..874V, 2007A&A...464L..57S, 
2008MNRAS.389.1828S, 2008ApJ...677..556T, 
2009ApJ...704.1262T, 2011ApJ...728L..30G,
2011ApJ...728L..29T, 2011A&A...533A.139W, 
2012ApJ...744..123T, 2012ApJ...753...49V,
2013ApJ...768...34B, 2014A&A...570A..58W,
2014ApJ...795..118Z, 2015A&A...584A.105D}.
In the astrophysical context the instability is refereed to as 
fingering convection, and it is well known to lead to 
efficient extra-mixing.

Preliminary exploration of the effect of the fingering convection in 
the case of static white dwarf models accreting  debris disk material 
has shown that ignoring this extra-mixing leads to underestimate the 
accretion rates by several orders of magnitude for DA 
(hydrogen-atmosphere) white dwarfs. By contrast it has negligible 
or marginal effect on the accretion rate on DB (helium-atmosphere) 
white dwarfs \citep{2013A&A...557L..12D}. 

In the present paper, we extend the previous study by simulating the time dependent 
accretion of debris-disk material on two DA white dwarfs. We selected the two DAZ G29-38, 
the prototype of this class, and GD133.
Both have an IR excess, signature of a  debris disk 
\citep{1987Natur.330..138Z, 2005ApJ...635L.161R,2007ApJ...663.1285J,
2009AJ....137.3191J,2009ApJ...694..805F},
 have a heavily polluted 
atmosphere and, in addition, both are ZZ Ceti pulsators. 
Their parameters and abundances are rather well 
determined from spectroscopy \citep{2014ApJ...783...79X}
 and asteroseismology 
\citep{2013RAA....13.1438C, 2016Fu_Vau_Su}.

The paper is organized as follows. Section 2 indicates how 
we obtain the initial models for the simulations. In section 
3 we describe the numerical simulations and their  
results are given in section 4. We summarize and discuss our 
results in section 5.

\section{Initial models}  
\label{sec:modini}

To  assess  the  impact  of  fingering  convection  on  the  estimated
accretion   rates,  we   concentrate  on   two  different   models  of
hydrogen-rich  (DA) white  dwarfs. Although  fingering  convection has
been  shown to  be  a  relevant physical  process  in accreting  white
dwarfs \citep{2013A&A...557L..12D},  the role  of this process  on the
expected accretion rates  in evolving white dwarf still remains  to be
investigated.  In this paper we  take this endeavour, by analysing two
white dwarfs,  GD133 and  G29-38. We based  our study on  a consistent
treatment of mixing processes and white dwarf evolution.  We performed
two  different  types  of  numerical experiments:  one  assuming  pure
gravitational   settling   and   another  including   also   fingering
convection.    Table  \ref{tab:params}   shows  the   adopted  stellar
parameters for both stars\footnote{For the particular case of 
G29-38 there is a large variety of parameters estimated by different 
authors \citep[see][]{2012MNRAS.420.1462R, 
2013RAA....13.1438C, 2014ApJ...783...79X}. We decided to
use the asteroseismological parameters of \cite{2013RAA....13.1438C} 
in order to study a more massive model and also to avoid 
the case of an unusually thin hydrogen envelope 
as that estimated by \cite{2012MNRAS.420.1462R}.}

%
%
\begin{table}

\caption{ Stellar parameters adopted for G29-38 and GD133. 
For G29-38, the mass and the effective temperature are the values 
derived from spectroscopy \citep{2014ApJ...783...79X}. For GD133, 
the values are derived from the asteroseismology analysis of 
\cite{2016Fu_Vau_Su}. They are in good agreement with the values 
given by spectroscopy \citep{2014ApJ...783...79X}. The hydrogen mass 
fractions are from asteroseismology: \cite{2013RAA....13.1438C} 
for G~29-38, and \cite{2016Fu_Vau_Su} for GD~133.}
\label{tab:params}
\centering
\renewcommand{\arraystretch}{1.2}
\begin{tabular}{ccccc}
\hline\hline
 star & $M_\mathrm{WD}$ [$M_\odot$] & $T_\mathrm{eff}$ [K] &
 $\log M_\mathrm{H}/M_\mathrm{WD}$ &  $\log M_\mathrm{cvz}/M_\mathrm{WD}$ \\
\hline
G29-38 & $0.85$ & $11820 \pm 100$ & $ 10^{-4}$ & $-13.95$ \\
GD133  & $0.63$ & $12400 \pm 100$ & $3.16 \times 10^{-5}$ & $-16.19$ \\
\hline
\end{tabular}
\end{table}

Our     initial    white    dwarf     models    were     taken    from
\cite{2010ApJ...717..183R}.  These  models  are  the  result  of  full
evolutionary calculations from the zero age main sequence, through the
core  hydrogen  burning  phase,  the  helium  burning  phase  and  the
thermally  pulsing asymptotic giant  branch phase  to the  white dwarf
stage.  Specifically,  we  selected  two models  with  stellar  masses
closest  to  that of  our  objects  of interest:  $M_\mathrm{WD}=0.632
M_\odot$ ($Z=0.01$, M=$2.25 M_\odot$ progenitor),  for  GD133, and  
$M_\mathrm{WD}=0.837  M_\odot$
($Z=0.01$,  M=$4.0 M_\odot$ progenitor), for G29-38.  
These  models were evolved from the beginning
of  the  cooling  branch   to  lower  luminosities,  allowing  element
diffusion  to   operate.   As  soon  as  heavier   elements  sunk,  we
artificially modified the envelope of the models in order to match the
amount  of hydrogen  according to  the  asteroseismological inferences
\citep{2013RAA....13.1438C, 2016Fu_Vau_Su}.
We evolved the models until they reached the 
final effective temperature listed in Table \ref{tab:params}, where we 
stopped the simulations. The resulting models are considered   
as our {\it initial models}. From this stage, accretion is allowed to operate.

\section{Numerical computations}  
\label{simulations}

All computations were done using the {\tt LPCODE} stellar evolutionary
code \citep{2005A&A...435..631A,
  2013A&A...557A..19A,2015A&A...576A...9A}, properly modified in order
to simulate the accretion scenario and to follow the evolution of some
chemical elements that were not present in the original version of the
code. These chemical elements
are  $^{24}$Mg,  $^{28}$Si,  $^{40}$Ca,  and $^{56}$Fe,  with  initial
abundances in solar proportions. 
Gravitational settling was included
following Burger's scheme \citep{1969fecg.book.....B}.
For details about our stellar code, and
the input physics, we refer the reader to the above cited papers.

There has been some discussions about whether the accretion from the 
debris disk occurs as a continuous or a discontinuous process and whether
the accreted material follows a particular geometry. In the case of G~29-38,
\cite{2010ApJ...714..296T}
find a hint that the accretion geometry could be
time dependent, pointing to a discontinuous accretion producing spots 
of heavy elements on the white dwarf surface \citep[see also][]{2008ApJ...685L.133M}
and to the influence of a potential magnetic field on the accretion process. 
However, the time variability of the heavy element abundances,  Ca in the 
case of G~29-38, which 
would be the signature of such a discontinuous accretion, 
was not confirmed \citep{2008ApJ...677L..43D}. 
In addition GD ~133 and G~29-38 have no detected magnetic field \citep{2015SSRv..191..111F}. 

In the present simulations, accretion  was  implemented  as  a continuous  process
and the accreted material is assumed to be uniformly
distributed on the star's surface.  We  performed
simulations for different accretion  rates, ranging from $10^4$ g/s to
$10^{10}$ g/s.  Metal abundances of  accreted matter were set  to mimic
the  composition of  the bulk  Earth  \citep{2001E&PSL.185...49A}. The
amount of  accreted matter is  mixed instantaneously in  the convective
zone, by assuming  that the mixing timescale is  much shorter than the
characteristic cooling timescale.

Our computations were done in the framework of the double diffusion theory
(GNA theory) developed by \cite{1993ApJ...407..284G}, adopting for the 
diffusion coefficient of fingering convection zones the prescription of
\cite{2013ApJ...768...34B}. This formulation allows the treatment  of the fingering convection instability 
 \citep[for more details,   see][]{2011A&A...533A.139W}. We also  performed computations using
the  classical   convection  theory  (ML2,   with  $\alpha=0.8$).  Whereas fingering convection is included 
in the GNA computations, it is not taken into account in the simple MLT method. For
numerical  purposes,  small  evolutionary  time  steps  were  adopted,
particularly  for  large accretion rates.  Specifically,  we used
time steps of the orders of  days for ML2 simulations and of the order
of {\it hours}, for GNA experiments.

\begin{figure}
\resizebox{\hsize}{!}{\includegraphics{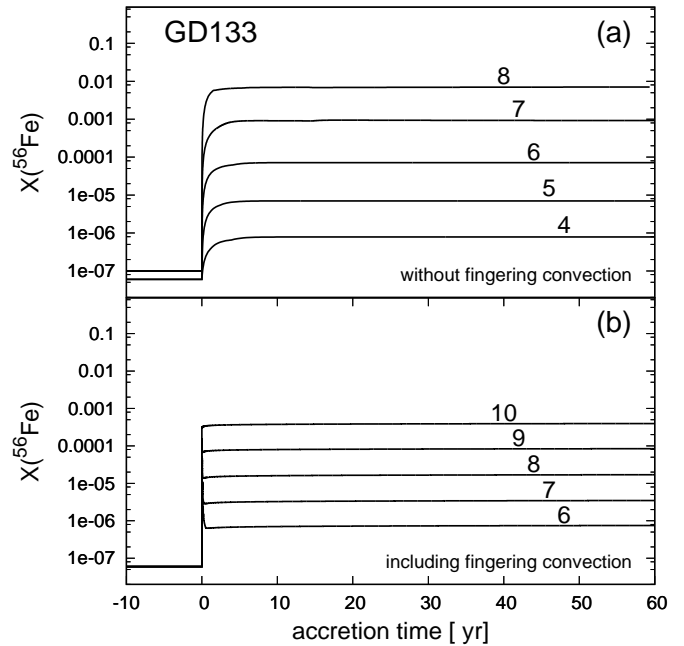}}  
\caption{Time evolution of $^{56}$Fe surface abundance for the model of GD133. 
{\it Top panel:} without taking fingering
convection into account. {\it Bottom panel:} including fingering convection.
Numbers indicate the logarithm of the adopted accretion rates, in units of
g/s.}
\label{fig:Fe56-GD133}
\end{figure}

\section{Results}  
\label{results}

Time evolution of surface  composition for the model representative of
GD133,  once  accretion  has  been  turned on,  is  displayed  in  Fig
\ref{fig:Fe56-GD133} for  the various adopted values  of the accretion
rate.   For the  sake of  clarity we  only show  one  element, namely,
$^{56}$Fe.  The  situation  for  other heavy  elements  is  similar,
although each one  reaches its own final abundance  level according to
its relative  presence in  the accreted matter  and to  its individual
diffusion velocity.  From Fig.  \ref{fig:Fe56-GD133} we see  that, for
the same accretion rate, the surface contamination becomes larger when
fingering  convection is  disregarded.  This is  because the  accreted
material  mixes almost instantaneously  in the  convective zone,  thus
leading  to the formation  of a  chemical inversion  (heavier material
ontop  lighter one) at  its base.  In the  framework of  the canonical
convection theory  (MLT), this chemical  inversion does not  alter the
chemical  structure in deeper layers; there,  abundance changes are
only due to gravitational settling.   In contrast, when the GNA convection 
theory
is  considered,  the chemical  inversion  triggers  the appearance  of
fingering convection  below the  convective zone.   As a
consequence, heavy elements become  more efficiently mixed into deeper
layers,  reducing their  abundances in  the atmosphere  (mainly  by {\it
  dilution}).  Diffusion  may  also  contribute  to  lower  the  metal
contamination of  the atmosphere but, as compared  with the efficiency
of fingering  convection, its contribution  is much smaller.  Fig. \ref{fig:Fe56-GD133} clearly  illustrates that a
stationary  state is  rapidly reached  for all  the  adopted accretion
rates. In fact, less than 10 yr of accretion is needed for the surface
abundances to be stabilized. In the  case of the MLT, shorter times are
needed to reach a stationary state.

\begin{figure}
\resizebox{\hsize}{!}{\includegraphics{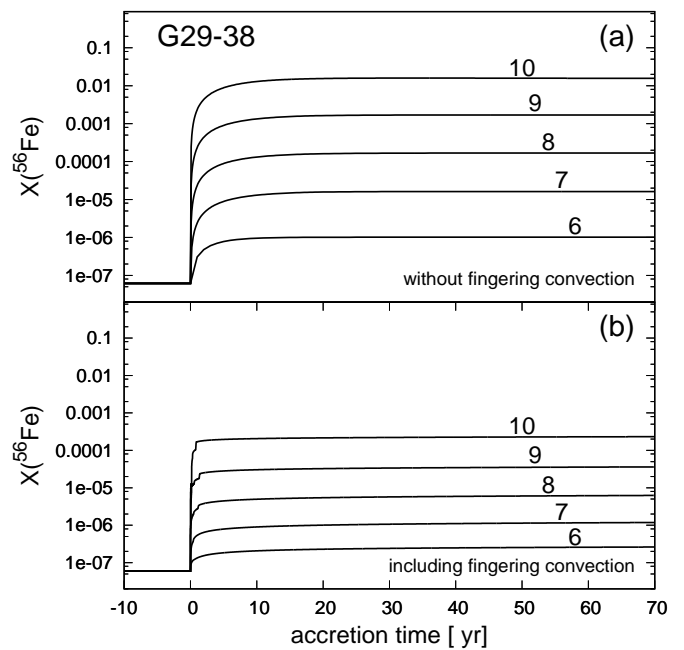}}  
\caption{Same as Fig. \ref{fig:Fe56-GD133} for G29-38.}
\label{fig:Fe56-G29-38}
\end{figure}

The results for G29-38 are depicted in Fig.\ref{fig:Fe56-G29-38}. 
As compared with the situation in GD133, longer times are needed
to reach the stationary state. Note also  that, when  fingering
convection is not considered,  much larger 
accretion rates  (almost two orders
of magnitude)  are needed in G29-38 than in GD133
to reach a given surface contamination.
This is related to the depth reached by the convective 
region, which is about a hundred times more massive
for G29-38 than for GD133, as can be seen in Table 
\ref{tab:params}. This is different when fingering convection is allowed, because it mixes matter deeper inside the star. 
\begin{figure}
\resizebox{\hsize}{!}{\includegraphics{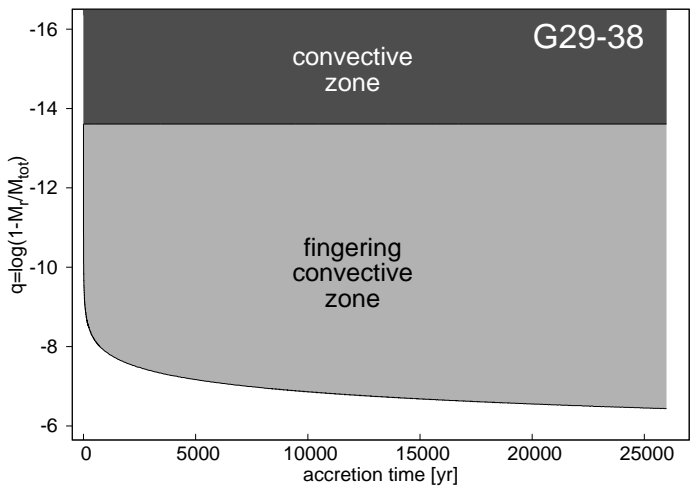}}  
\caption{Time evolution of the stellar structure for the
G29-38 model. The accretion rate was of $10^{10}$ g/s.}
\label{fig:fcz-t}
\end{figure}

Fig.\ref{fig:fcz-t} shows the behavior of fingering convection in the presence
of  accretion for our  maximum accretion  rate ($10^{10}$  g/s) for G~29-38.
 Note
that the fingering  convection zone grows very rapidly  once accretion has
started. In  a few  centuries, it spans  over a region  which embraces
about 1000  times more mass  than the convective zone.   This strongly
dilutes heavy elements, lowering  surface contamination. It is kind of
surprising  that  the fingering  convective  zone  was still  evolving
deeper after more than 25000  years of continuous accretion.  It would
have been interesting  to follow this experiment until  a steady state
could  be reached  but unfortunately  we  had to  stop the  simulation
because    of   the   (prohibitively)    long   run    time   required
(Fig. \ref{fig:fcz-t} demanded one month of cpu time). In any case, it
is  evident that  the fingering convection  zone goes on  growing  for a
period of time substantially much longer than the time needed for surface
elements to reach a stationary  state.  A similar situation was also observed
for the GD133 case. The whole set of simulations shows that the higher
the accretion rate the larger the fingering convection region. In this
sense, Fig. \ref{fig:fcz-t} illustrates the situation of maximum depth
reached by this zone in our simulations.

\begin{figure}
\resizebox{\hsize}{!}{\includegraphics{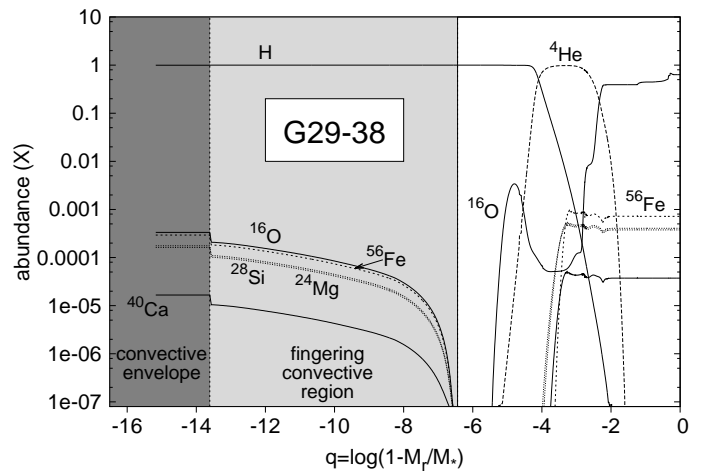}}  
\caption{Final chemical profile for our G29-38 model submitted
to an accretion rate of $10^{10}$ g/s. The lines corresponding to 
$^{24}$Mg and $^{28}$Si almost overlap. The same happens to 
$^{16}$O and $^{56}$Fe. The bottom of the fingering convection
zone does never reach central heavy elements.}
\label{fig:chemprof-G29-38}
\end{figure}

Fig. \ref{fig:chemprof-G29-38} shows  the internal chemical profiles of
the  last  computed  model  in  Fig. \ref{fig:fcz-t}  for  H,  $^4$He,
$^{16}$O,  $^{24}$Mg, $^{28}$Si, $^{40}$Ca,  and $^{56}$Fe.  The dark grey
region marks the dynamical convective envelope,  in which the elements are almost instantaneously mixed and thus distributed
homogeneously. Below this zone, there is a very extense fingering
convection region (light grey),  where heavy elements are carried down to  the   interior layers by  milder mixing, whose efficiency 
decreases toward the  bottom of  the fingering
convective region.  For this reason, this region  shows chemical  profiles where
heavy  elements abundances decrease with depth. Note that  the fingering  convective zone  never
reaches the compositional gradients inside the stellar core. Such a situation can occur however in  the case of a  very small hydrogen  envelope.  Since the
maximum depth reached by the fingering convective zone is proportional
to      the     accretion      rate      and     considering      that
Fig.  \ref{fig:chemprof-G29-38} shows  the situation  for  the largest
accretion rate,  it is clear that  none of our simulations  was able to
lead to element dredge-up from the deeper regions.

\begin{figure}
\resizebox{\hsize}{!}{\includegraphics{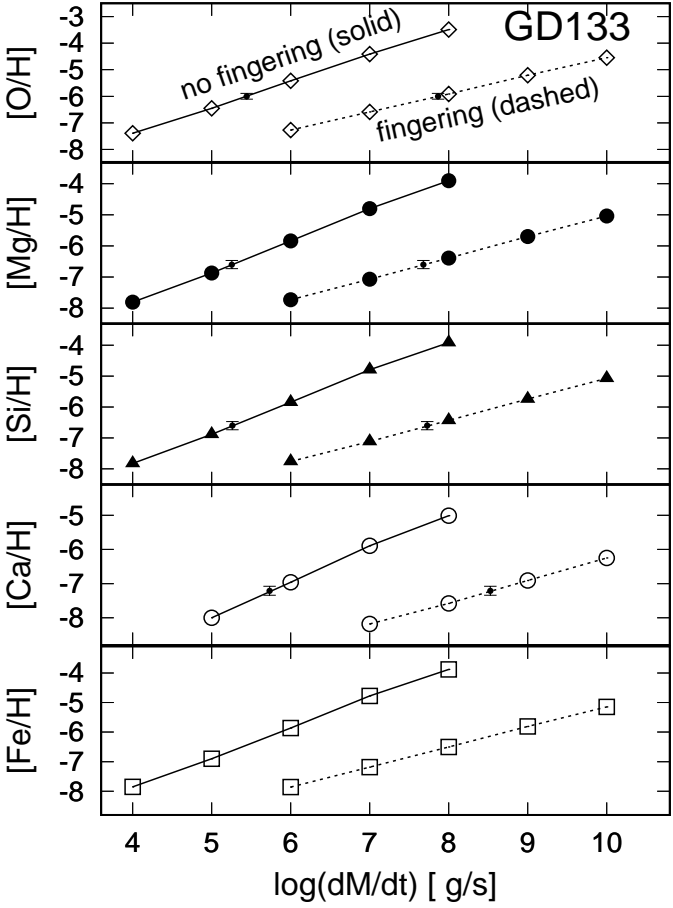}} 
\caption{Dependence of final abundances on the accretion rates and 
on the adopted convection theory for the case of GD133. 
Solid lines stand for results obtained 
by using the classical mixing length convection theory (MLT). Dashed 
lines connect points obtained by implementing a double diffusion 
convection theory (GNA). Points with error bars refer to the
observed abundances reported by \cite{2014ApJ...783...79X}.
For GD133, Fe was not detected by \cite{2014ApJ...783...79X}, 
thus no point was added in the corresponding figure.
$[X/H] = \log n(X)/n(H)$, the logarithmic number ratio of the 
abundance of element X relative to the abundance of H.}
\label{fig:abund-GD133}
\end{figure}

Fig. \ref{fig:abund-GD133} summarizes the final outcome of
the simulations 
for our GD133 model. Abundances are given as in 
\cite{2014ApJ...783...79X} for the purpose of comparison. 
The figure shows the final (steady state) surface
abundances obtained by our simulations as a function of the accretion
rate. Each chemical element is represented by a different symbol. Lines 
help us identifying the general behavior of the abundances as we change the
accretion rate. Dashed (solid)  lines stand for the situation when
fingering convection is considered (disregarded).
From Fig. \ref{fig:abund-GD133} 
we see that there is an almost linear relation between 
[X/H] and the logarithm of accretion rates. The abundances grow faster 
when fingering convection is not taken into account, as shown by the
different slopes of solid and dashed lines. 
We added to Fig. \ref{fig:abund-GD133}
some points (with error bars) corresponding to the observed
abundances of a few elements reported by \cite{2014ApJ...783...79X}
for GD133, and overlap them with the lines of the corresponding element.
If fingering convection is not considered, the observed points group
themselves around $\log(dM/dt)\approx 5.5$ (for [$dM/dt$] in g/s)
while if  fingering convection is taken into account, points are
compatible with a much larger accretion rate, roughly around 
$\log(dM/dt)\approx 8$. Hence, neglecting fingering convection 
may lead to a difference of more than 2 orders of
magnitude when estimating the accretion rate for the case of GD133.

\begin{figure}
\resizebox{\hsize}{!}{\includegraphics{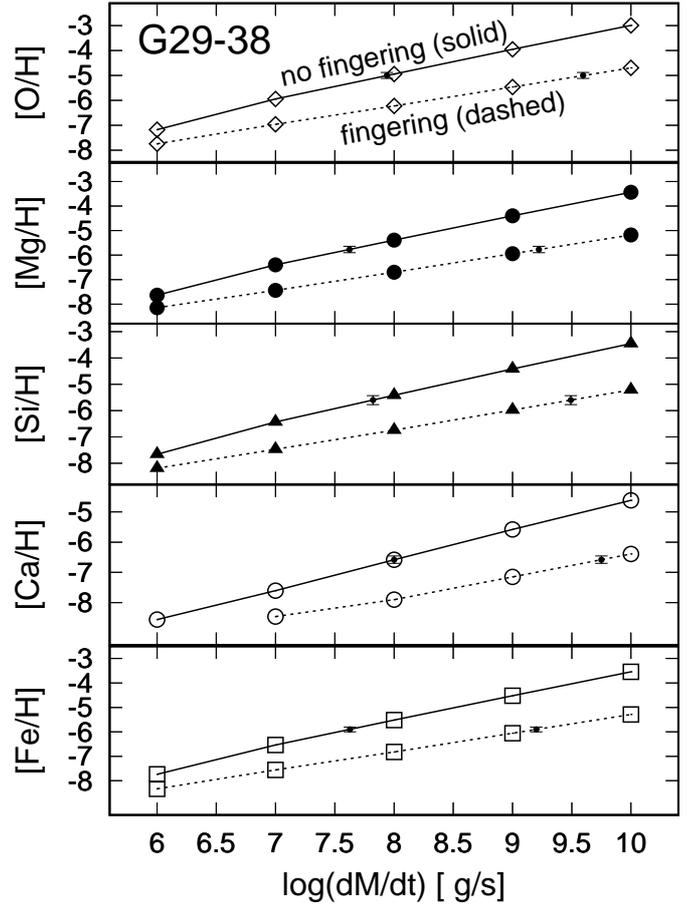}} 
\caption{Same as Fig. \ref{fig:abund-GD133} but for G29-38.}
\label{fig:abund-G29-38}
\end{figure}

Fig. \ref{fig:abund-G29-38} shows the situation for  G29-38. Some
differences are evident. Perhaps the most conspicuous one
is that the dependence of the final abundance on the accretion rate
is less steep here than in GD133. This can be understood on the basis that 
G29-38 has a convective zone that is 100 times larger 
(in mass) than GD133. This introduces a large difference in the resulting 
contamination between the two stars when fingering convection is not considered.
On the contrary, there is almost no such difference when
fingering convection is considered
(compare dashed lines of Figs. \ref{fig:abund-GD133}
and \ref{fig:abund-G29-38}). The final surface abundances 
depend mainly on the total size of the envelope's mixed 
region and not on the original size of the convective zone.

Figure \ref{fig:abund-G29-38} also includes a few points with
error bars, corresponding to the abundances reported by 
\cite{2014ApJ...783...79X} for G29-38. We see that 
according to our ``no fingering'' simulations, the
observed abundances should result from an accretion process
at a rate of about $\log(dM/dt)\approx 7.8$. In contrast, 
numerical experiments including fingering convection
suggest that the accretion rate should be larger, around 
$\log(dM/dt)\approx 9.5$. Thus, for G29-38 we see that
the difference between both
estimates is smaller than for GD 133, namely, of approximately $1.7$ dex.
In any case, we find that the inclusion of fingering convection
implies much larger accretion rates to explain observed abundances than
inferences based on the standard MLT.

\section{Summary and Discussion}
\label{discussion}
We have performed numerical simulations of a continuous and 
uniform accretion of material with chemical composition similar 
to the bulk Earth composition onto the two DAZ and ZZ Ceti 
pulsators GD~133 and G~29-38. These two stars have 
been selected because they are both surrounded by a 
debris disk revealed by their infrared excess, have a detailed 
heavy element abundance analysis and have well constrained 
global parameters, including their hydrogen mass-fraction, from 
asteroseismology. 

Accretion rates ranging from $10^4$ g/s to $10^{10}$ g/s have 
been used in these simulations. Two cases have been considered: 
in the first case the accreted material was mixed in the surface 
convection zone, described by the standard mixing-length 
theory (MLT), and diffused through its bottom by gravitational 
settling, while in the second case we introduced in addition the double-diffusive 
instability induced by the building of the inverse  $\mu$-gradient resulting from the accretion of heavy elements 
on top of the hydrogen-rich outer layers. In this second case, 
the instability is efficient in producing an extra-mixing by 
fingering convection. 
As a consequence, the accreted material is mixed in a much deeper region of the star. 
The resulting surface abundances of the heavy elements which have been considered 
in this study (Mg, O, Ca, Fe, Si) very rapidly reach a stationary value
while the fingering convection regularly progress downwards. As a result of the 
extra-mixing the accretion rates needed to reproduce the abundances derived from the 
observations significantly exceed the rates estimated when the fingering-convection 
is ignored. In the cases of GD~133, we find that the accretion rate
needed to reproduce the observed abundances is $\log(dM/dt)\approx 5.5$ (for [$dM/dt$] in g/s)
when fingering convection is not taken into account, but rises to $\log(dM/dt)\approx 8$
when fingering convection is properly taken into account. For G~29-38 the observed 
abundances should result from an accretion rate $\log(dM/dt)\approx 7.8$ according 
to our ``no fingering'' simulations but an accretion rate around $\log(dM/dt)\approx 9.5$ 
is needed when fingering convection is included.

We conclude that the extra-mixing induced by the fingering convection is an important 
effect which has to be taken into account when estimating the accretion rates needed to reproduce
the heavy element abundances observed in the DAZ white dwarfs. It may imply that 
significantly more massive rocky bodies may have been disrupted by tidal effects than previously estimated.

\begin{acknowledgements}
Part of  this work was  supported by the Agencia Nacional de Promoci\'on Cient\'{\i}fica y Tecnol\'ogica (ANPCyT) through the Programa de
Modernizaci\'on Tecnol\'gica BID 1728/OC-AR, by the PIP
112-200801-00940 grant from CONICET (Argentina). 
Some simulations were run on IFLySiB's Cluster, which is part of 
SNCAD-MinCyT, Argentina.
\end{acknowledgements}  
  
\bibliographystyle{aa} 
\bibliography{twoDAs} 

\begin{thebibliography}{57}
\expandafter\ifx\csname natexlab\endcsname\relax\def\natexlab#1{#1}\fi

\bibitem[{{All{\`e}gre} {et~al.}(2001){All{\`e}gre}, {Manh{\`e}s}, \&
  {Lewin}}]{2001E&PSL.185...49A}
{All{\`e}gre}, C., {Manh{\`e}s}, G., \& {Lewin}, {\'E}. 2001, Earth and
  Planetary Science Letters, 185, 49

\bibitem[{{Althaus} {et~al.}(2015){Althaus}, {Camisassa}, {Miller Bertolami},
  {C{\'o}rsico}, \& {Garc{\'{\i}}a-Berro}}]{2015A&A...576A...9A}
{Althaus}, L.~G., {Camisassa}, M.~E., {Miller Bertolami}, M.~M., {C{\'o}rsico},
  A.~H., \& {Garc{\'{\i}}a-Berro}, E. 2015, \aap, 576, A9

\bibitem[{{Althaus} {et~al.}(2013){Althaus}, {Miller Bertolami}, \&
  {C{\'o}rsico}}]{2013A&A...557A..19A}
{Althaus}, L.~G., {Miller Bertolami}, M.~M., \& {C{\'o}rsico}, A.~H. 2013,
  \aap, 557, A19

\bibitem[{{Althaus} {et~al.}(2005){Althaus}, {Serenelli}, {Panei},
  {C{\'o}rsico}, {Garc{\'{\i}}a-Berro}, \&
  {Sc{\'o}ccola}}]{2005A&A...435..631A}
{Althaus}, L.~G., {Serenelli}, A.~M., {Panei}, J.~A., {et~al.} 2005, \aap, 435,
  631

\bibitem[{{Brown} {et~al.}(2013){Brown}, {Garaud}, \&
  {Stellmach}}]{2013ApJ...768...34B}
{Brown}, J.~M., {Garaud}, P., \& {Stellmach}, S. 2013, \apj, 768, 34

\bibitem[{{Burgers}(1969)}]{1969fecg.book.....B}
{Burgers}, J.~M. 1969, {Flow Equations for Composite Gases}

\bibitem[{{Chen} \& {Li}(2013)}]{2013RAA....13.1438C}
{Chen}, Y.-H. \& {Li}, Y. 2013, Research in Astronomy and Astrophysics, 13,
  1438

\bibitem[{{Croll} {et~al.}(2015){Croll}, {Dalba}, {Vanderburg}, {Eastman},
  {Rappaport}, {DeVore}, {Bieryla}, {Muirhead}, {Han}, {Latham}, {Beatty},
  {Wittenmyer}, {Wright}, {Johnson}, \& {McCrady}}]{2015arXiv151006434C}
{Croll}, B., {Dalba}, P.~A., {Vanderburg}, A., {et~al.} 2015, ArXiv e-prints

\bibitem[{{Deal} {et~al.}(2013){Deal}, {Deheuvels}, {Vauclair}, {Vauclair}, \&
  {Wachlin}}]{2013A&A...557L..12D}
{Deal}, M., {Deheuvels}, S., {Vauclair}, G., {Vauclair}, S., \& {Wachlin},
  F.~C. 2013, \aap, 557, L12

\bibitem[{{Deal} {et~al.}(2015){Deal}, {Richard}, \&
  {Vauclair}}]{2015A&A...584A.105D}
{Deal}, M., {Richard}, O., \& {Vauclair}, S. 2015, \aap, 584, A105

\bibitem[{{Debes} \& {L{\'o}pez-Morales}(2008)}]{2008ApJ...677L..43D}
{Debes}, J.~H. \& {L{\'o}pez-Morales}, M. 2008, \apjl, 677, L43

\bibitem[{{Debes} \& {Sigurdsson}(2002)}]{2002ApJ...572..556D}
{Debes}, J.~H. \& {Sigurdsson}, S. 2002, \apj, 572, 556

\bibitem[{{Debes} {et~al.}(2012){Debes}, {Walsh}, \&
  {Stark}}]{2012ApJ...747..148D}
{Debes}, J.~H., {Walsh}, K.~J., \& {Stark}, C. 2012, \apj, 747, 148

\bibitem[{{Farihi} {et~al.}(2012){Farihi}, {G{\"a}nsicke}, {Wyatt}, {Girven},
  {Pringle}, \& {King}}]{2012MNRAS.424..464F}
{Farihi}, J., {G{\"a}nsicke}, B.~T., {Wyatt}, M.~C., {et~al.} 2012, \mnras,
  424, 464

\bibitem[{{Farihi} {et~al.}(2009){Farihi}, {Jura}, \&
  {Zuckerman}}]{2009ApJ...694..805F}
{Farihi}, J., {Jura}, M., \& {Zuckerman}, B. 2009, \apj, 694, 805

\bibitem[{{Ferrario} {et~al.}(2015){Ferrario}, {de Martino}, \&
  {G{\"a}nsicke}}]{2015SSRv..191..111F}
{Ferrario}, L., {de Martino}, D., \& {G{\"a}nsicke}, B.~T. 2015, \ssr, 191, 111

\bibitem[{{Frewen} \& {Hansen}(2014)}]{2014MNRAS.439.2442F}
{Frewen}, S.~F.~N. \& {Hansen}, B.~M.~S. 2014, \mnras, 439, 2442

\bibitem[{{Fu} {et~al.}(2016){Fu}, {Vauclair}, {Su}, {Cang}, {Li}, {Xue},
  {Jiang}, {Li}, {Niu}, {Zhang}, {Colas}, {Dolez}, {Fox Machado}, {Michel},
  {Alvarez}, \& {Kim}}]{2016Fu_Vau_Su}
{Fu}, J.-N., {Vauclair}, G., {Su}, J., {et~al.} 2016, 20th European Workshop on
  White Dwarfs, (in press)

\bibitem[{{G{\"a}nsicke} {et~al.}(2016){G{\"a}nsicke}, {Aungwerojwit}, {Marsh},
  {Dhillon}, {Sahman}, {Veras}, {Farihi}, {Chote}, {Ashley}, {Arjyotha},
  {Rattanasoon}, {Littlefair}, {Pollacco}, \& {Burleigh}}]{2016ApJ...818L...7G}
{G{\"a}nsicke}, B.~T., {Aungwerojwit}, A., {Marsh}, T.~R., {et~al.} 2016,
  \apjl, 818, L7

\bibitem[{{Garaud}(2011)}]{2011ApJ...728L..30G}
{Garaud}, P. 2011, \apjl, 728, L30

\bibitem[{{Grossman} {et~al.}(1993){Grossman}, {Narayan}, \&
  {Arnett}}]{1993ApJ...407..284G}
{Grossman}, S.~A., {Narayan}, R., \& {Arnett}, D. 1993, \apj, 407, 284

\bibitem[{{Jura}(2003)}]{2003ApJ...584L..91J}
{Jura}, M. 2003, \apjl, 584, L91

\bibitem[{{Jura} {et~al.}(2007){Jura}, {Farihi}, \&
  {Zuckerman}}]{2007ApJ...663.1285J}
{Jura}, M., {Farihi}, J., \& {Zuckerman}, B. 2007, \apj, 663, 1285

\bibitem[{{Jura} {et~al.}(2009){Jura}, {Farihi}, \&
  {Zuckerman}}]{2009AJ....137.3191J}
{Jura}, M., {Farihi}, J., \& {Zuckerman}, B. 2009, \aj, 137, 3191

\bibitem[{{Jura} \& {Young}(2014)}]{2014AREPS..42...45J}
{Jura}, M. \& {Young}, E.~D. 2014, Annual Review of Earth and Planetary
  Sciences, 42, 45

\bibitem[{{Koester} {et~al.}(2014){Koester}, {G{\"a}nsicke}, \&
  {Farihi}}]{2014A&A...566A..34K}
{Koester}, D., {G{\"a}nsicke}, B.~T., \& {Farihi}, J. 2014, \aap, 566, A34

\bibitem[{{Melis} {et~al.}(2011){Melis}, {Farihi}, {Dufour}, {Zuckerman},
  {Burgasser}, {Bergeron}, {Bochanski}, \& {Simcoe}}]{2011ApJ...732...90M}
{Melis}, C., {Farihi}, J., {Dufour}, P., {et~al.} 2011, \apj, 732, 90

\bibitem[{{Montgomery} {et~al.}(2008){Montgomery}, {Thompson}, \& {von
  Hippel}}]{2008ApJ...685L.133M}
{Montgomery}, M.~H., {Thompson}, S.~E., \& {von Hippel}, T. 2008, \apjl, 685,
  L133

\bibitem[{{Mustill} {et~al.}(2013){Mustill}, {Villaver}, {Veras}, {Bonsor}, \&
  {Wyatt}}]{2013EPJWC..4706008M}
{Mustill}, A.~J., {Villaver}, E., {Veras}, D., {Bonsor}, A., \& {Wyatt}, M.~C.
  2013, in European Physical Journal Web of Conferences, Vol.~47, European
  Physical Journal Web of Conferences, 06008

\bibitem[{{Reach} {et~al.}(2005){Reach}, {Kuchner}, {von Hippel}, {Burrows},
  {Mullally}, {Kilic}, \& {Winget}}]{2005ApJ...635L.161R}
{Reach}, W.~T., {Kuchner}, M.~J., {von Hippel}, T., {et~al.} 2005, \apjl, 635,
  L161

\bibitem[{{Renedo} {et~al.}(2010){Renedo}, {Althaus}, {Miller Bertolami},
  {Romero}, {C{\'o}rsico}, {Rohrmann}, \&
  {Garc{\'{\i}}a-Berro}}]{2010ApJ...717..183R}
{Renedo}, I., {Althaus}, L.~G., {Miller Bertolami}, M.~M., {et~al.} 2010, \apj,
  717, 183

\bibitem[{{Romero} {et~al.}(2012){Romero}, {C{\'o}rsico}, {Althaus}, {Kepler},
  {Castanheira}, \& {Miller Bertolami}}]{2012MNRAS.420.1462R}
{Romero}, A.~D., {C{\'o}rsico}, A.~H., {Althaus}, L.~G., {et~al.} 2012, \mnras,
  420, 1462

\bibitem[{{Stancliffe} \& {Glebbeek}(2008)}]{2008MNRAS.389.1828S}
{Stancliffe}, R.~J. \& {Glebbeek}, E. 2008, \mnras, 389, 1828

\bibitem[{{Stancliffe} {et~al.}(2007){Stancliffe}, {Glebbeek}, {Izzard}, \&
  {Pols}}]{2007A&A...464L..57S}
{Stancliffe}, R.~J., {Glebbeek}, E., {Izzard}, R.~G., \& {Pols}, O.~R. 2007,
  \aap, 464, L57

\bibitem[{{Th{\'e}ado} \& {Vauclair}(2012)}]{2012ApJ...744..123T}
{Th{\'e}ado}, S. \& {Vauclair}, S. 2012, \apj, 744, 123

\bibitem[{{Th{\'e}ado} {et~al.}(2009){Th{\'e}ado}, {Vauclair}, {Alecian}, \&
  {LeBlanc}}]{2009ApJ...704.1262T}
{Th{\'e}ado}, S., {Vauclair}, S., {Alecian}, G., \& {LeBlanc}, F. 2009, \apj,
  704, 1262

\bibitem[{{Thompson} {et~al.}(2008){Thompson}, {Ivans}, {Bisterzo}, {Sneden},
  {Gallino}, {Vauclair}, {Burley}, {Shectman}, \&
  {Preston}}]{2008ApJ...677..556T}
{Thompson}, I.~B., {Ivans}, I.~I., {Bisterzo}, S., {et~al.} 2008, \apj, 677,
  556

\bibitem[{{Thompson} {et~al.}(2010){Thompson}, {Montgomery}, {von Hippel},
  {Nitta}, {Dalessio}, {Provencal}, {Strickland}, {Holtzman}, {Mukadam},
  {Sullivan}, {Nagel}, {Koziel-Wierzbowska}, {Kundera}, {Zola}, {Winiarski},
  {Drozdz}, {Kuligowska}, {Ogloza}, {Bogn{\'a}r}, {Handler}, {Kanaan},
  {Ribeira}, {Rosen}, {Reichart}, {Haislip}, {Barlow}, {Dunlap}, {Ivarsen},
  {LaCluyze}, \& {Mullally}}]{2010ApJ...714..296T}
{Thompson}, S.~E., {Montgomery}, M.~H., {von Hippel}, T., {et~al.} 2010, \apj,
  714, 296

\bibitem[{{Traxler} {et~al.}(2011){Traxler}, {Garaud}, \&
  {Stellmach}}]{2011ApJ...728L..29T}
{Traxler}, A., {Garaud}, P., \& {Stellmach}, S. 2011, \apjl, 728, L29

\bibitem[{{Vanderburg} {et~al.}(2015){Vanderburg}, {Johnson}, {Rappaport},
  {Bieryla}, {Irwin}, {Lewis}, {Kipping}, {Brown}, {Dufour}, {Ciardi}, {Angus},
  {Schaefer}, {Latham}, {Charbonneau}, {Beichman}, {Eastman}, {McCrady},
  {Wittenmyer}, \& {Wright}}]{2015Natur.526..546V}
{Vanderburg}, A., {Johnson}, J.~A., {Rappaport}, S., {et~al.} 2015, \nat, 526,
  546

\bibitem[{{Vauclair}(2004)}]{2004ApJ...605..874V}
{Vauclair}, S. 2004, \apj, 605, 874

\bibitem[{{Vauclair} \& {Th{\'e}ado}(2012)}]{2012ApJ...753...49V}
{Vauclair}, S. \& {Th{\'e}ado}, S. 2012, \apj, 753, 49

\bibitem[{{Veras}(2016)}]{2016RSOS....3.0571V}
{Veras}, D. 2016, Royal Society Open Science, 3, 150571

\bibitem[{{Veras} {et~al.}(2015{\natexlab{a}}){Veras}, {Eggl}, \&
  {G{\"a}nsicke}}]{2015MNRAS.452.1945V}
{Veras}, D., {Eggl}, S., \& {G{\"a}nsicke}, B.~T. 2015{\natexlab{a}}, \mnras,
  452, 1945

\bibitem[{{Veras} {et~al.}(2015{\natexlab{b}}){Veras}, {Eggl}, \&
  {G{\"a}nsicke}}]{2015MNRAS.451.2814V}
{Veras}, D., {Eggl}, S., \& {G{\"a}nsicke}, B.~T. 2015{\natexlab{b}}, \mnras,
  451, 2814

\bibitem[{{Veras} {et~al.}(2014{\natexlab{a}}){Veras}, {Jacobson}, \&
  {G{\"a}nsicke}}]{2014MNRAS.445.2794V}
{Veras}, D., {Jacobson}, S.~A., \& {G{\"a}nsicke}, B.~T. 2014{\natexlab{a}},
  \mnras, 445, 2794

\bibitem[{{Veras} {et~al.}(2014{\natexlab{b}}){Veras}, {Leinhardt}, {Bonsor},
  \& {G{\"a}nsicke}}]{2014MNRAS.445.2244V}
{Veras}, D., {Leinhardt}, Z.~M., {Bonsor}, A., \& {G{\"a}nsicke}, B.~T.
  2014{\natexlab{b}}, \mnras, 445, 2244

\bibitem[{{Veras} {et~al.}(2015{\natexlab{c}}){Veras}, {Leinhardt}, {Eggl}, \&
  {G{\"a}nsicke}}]{2015MNRAS.451.3453V}
{Veras}, D., {Leinhardt}, Z.~M., {Eggl}, S., \& {G{\"a}nsicke}, B.~T.
  2015{\natexlab{c}}, \mnras, 451, 3453

\bibitem[{{Veras} {et~al.}(2013){Veras}, {Mustill}, {Bonsor}, \&
  {Wyatt}}]{2013MNRAS.431.1686V}
{Veras}, D., {Mustill}, A.~J., {Bonsor}, A., \& {Wyatt}, M.~C. 2013, \mnras,
  431, 1686

\bibitem[{{Veras} {et~al.}(2016){Veras}, {Mustill}, {G{\"a}nsicke}, {Redfield},
  {Georgakarakos}, {Bowler}, \& {Lloyd}}]{2016MNRAS.458.3942V}
{Veras}, D., {Mustill}, A.~J., {G{\"a}nsicke}, B.~T., {et~al.} 2016, \mnras,
  458, 3942

\bibitem[{{Wachlin} {et~al.}(2011){Wachlin}, {Miller Bertolami}, \&
  {Althaus}}]{2011A&A...533A.139W}
{Wachlin}, F.~C., {Miller Bertolami}, M.~M., \& {Althaus}, L.~G. 2011, \aap,
  533, A139

\bibitem[{{Wachlin} {et~al.}(2014){Wachlin}, {Vauclair}, \&
  {Althaus}}]{2014A&A...570A..58W}
{Wachlin}, F.~C., {Vauclair}, S., \& {Althaus}, L.~G. 2014, \aap, 570, A58

\bibitem[{{Xu} {et~al.}(2016){Xu}, {Jura}, {Dufour}, \&
  {Zuckerman}}]{2016ApJ...816L..22X}
{Xu}, S., {Jura}, M., {Dufour}, P., \& {Zuckerman}, B. 2016, \apjl, 816, L22

\bibitem[{{Xu} {et~al.}(2014){Xu}, {Jura}, {Koester}, {Klein}, \&
  {Zuckerman}}]{2014ApJ...783...79X}
{Xu}, S., {Jura}, M., {Koester}, D., {Klein}, B., \& {Zuckerman}, B. 2014,
  \apj, 783, 79

\bibitem[{{Zemskova} {et~al.}(2014){Zemskova}, {Garaud}, {Deal}, \&
  {Vauclair}}]{2014ApJ...795..118Z}
{Zemskova}, V., {Garaud}, P., {Deal}, M., \& {Vauclair}, S. 2014, \apj, 795,
  118

\bibitem[{{Zuckerman} \& {Becklin}(1987)}]{1987Natur.330..138Z}
{Zuckerman}, B. \& {Becklin}, E.~E. 1987, \nat, 330, 138

\bibitem[{{Zuckerman} {et~al.}(2010){Zuckerman}, {Melis}, {Klein}, {Koester},
  \& {Jura}}]{2010ApJ...722..725Z}
{Zuckerman}, B., {Melis}, C., {Klein}, B., {Koester}, D., \& {Jura}, M. 2010,
  \apj, 722, 725

\end{thebibliography}

\end{document}